\documentclass[11pt,a4paper]{report}

\usepackage[a4paper, margin=1.2in]{geometry}
\usepackage[utf8]{inputenc}
\usepackage{graphicx}
\usepackage[all]{xy}
\usepackage{amsmath}
\usepackage{amsfonts}
\usepackage{amssymb}
\usepackage{amsthm}
\usepackage{semantic}
\usepackage{mathtools}
\usepackage{array}
\usepackage{listings}
\usepackage{algorithm}
\usepackage[noend]{algpseudocode}
\usepackage{tikz}
\usepackage{multirow}
\usepackage{adjustbox}
\usepackage{pdflscape}
\usepackage{footnote}
\usepackage[font=footnotesize,labelfont=bf,margin=1cm]{caption}
\usepackage[%
    style=numeric, 
    sortcites,
    maxbibnames=99
]{biblatex}
\usepackage{url}

\newtheorem{definition}{Definition}

\newtheorem{example}[definition]{Example}

\let\olddefinition\definition
\renewcommand{\definition}{\olddefinition\normalfont}

\usetikzlibrary{arrows,positioning,shapes.geometric}

\makeatletter
\def\BState{\State\hskip-\ALG@thistlm}
\makeatother

\graphicspath{{imgs/}}

\begin{document}
\begin{titlepage}
\begin{center}
\textsc{\LARGE Research internship\\Computer Science}\\[1.5cm]
\includegraphics[height=100pt]{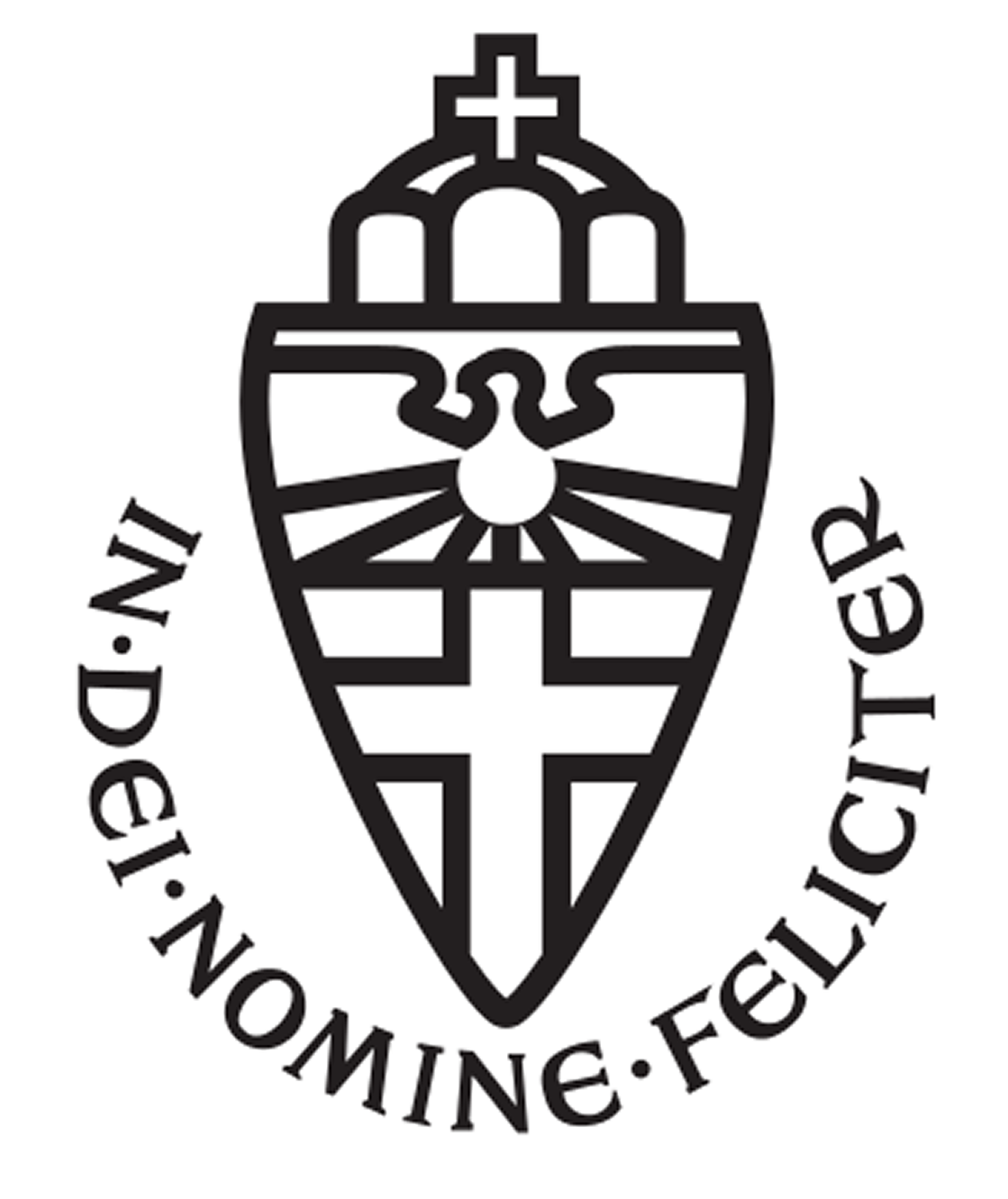}

\vspace{0.4cm}
\textsc{\Large Radboud University}\\[1cm]
\hrule
\vspace{0.4cm}
\textbf{\huge IdSan: An identity-based memory sanitizer for fuzzing binaries}\\[0.4cm]
\hrule
\vspace{2cm}
\begin{minipage}[t]{0.45\textwidth}
\begin{flushleft} \large
\textit{Author:}\\
Jos Craaijo\\
s4481674
\end{flushleft}
\end{minipage}
\begin{minipage}[t]{0.45\textwidth}
\begin{flushright} \large
\textit{First supervisor/assessor:}\\
Alexandru Geana\\
\texttt{geana@riscure.com}\\[1.3cm]
\textit{Second assessor:}\\
Erik Poll\\
\texttt{erik.poll@cs.ru.nl}
\end{flushright}
\end{minipage}
\vfill
{\large \today}
\end{center}
\end{titlepage}

This paper describes the results of an internship of 2 months at Riscure as part of my Master studies.

\begin{abstract}
    Most memory sanitizers work by instrumenting the program at compile time. There are only a handful of memory sanitizers that can sanitize a binary program without source code. Most are location-based, and are therefore unable to detect overflows of global variables or variables on the stack. In this paper we introduce an identity-based memory sanitizer for binary AArch64 programs which does not need access to the source code. It is able to detect overflows of stack- and global variables if the user provides some annotations or DWARF debugging information is available, as well as dynamically allocated memory.
\end{abstract}

\tableofcontents

\chapter{Introduction}

A sanitizer can help find bugs in programs that are being fuzzed. Most sanitizers require source code \cite{paricheck} \cite{bcc} \cite{bbc} \cite{asan} \cite{msan}. Memory sanitizers for binaries without source code are limited. QAsan \cite{qasan} and RetroWrite \cite{retrowrite}, both sanitizers for binaries, only work for dynamically allocated memory and entire stack frames, which leaves a large class of stack- or global overflows undetected. Especially for embedded devices, where dynamic memory allocation is much less common, the sanitizers currently available are lacking. MAI \cite{mai} is the most promising sanitizer for binaries, and attempts to automatically recover the full memory structure of a program. However it only supports x86-64 binaries, and has limitations like not being able to fully handle NULL-terminated strings, which presumably makes MAI less suitable for fuzzing large binaries.

In this paper we present IdSan, an identity-based memory sanitizer for fuzzing binaries. Because it is identity-based it can detect stack- and global overflows. It provides integrations with the unicornafl emulator and the Linux syscall \texttt{ptrace}. The interface to IdSan is minimal. Support for e.g. a debugging connection via JTAG or another processor architecture could be implemented in relatively little time. Rather than attempting to infer information on variable sizes at runtime like MAI, we use an approach where information on variable location and size is provided by a human. This allows for more accurate tracking of pointers, at the cost of a more time-consuming preparation. Information provided to IdSan does not need to be complete, and can be iteratively expanded.

% The design of IdSan is described in chapter \ref{design}. Next, in chapter \ref{implementation} we describe our implementation and trade-offs we made. In chapter \ref{evaluation} we evaluate our results. Finally, in chapter {ref} we conclude our paper.

\section{Fuzzing}
\emph{Fuzzing} is a commonly used technique to discover (security) bugs in software programs. A fuzzing tool (\emph{fuzzer}) repeatedly generates an input and provides this to the software program that is being fuzzed. It then observes the output of the program, and records interesting results (e.g. crashes or unexpected behavior) for human inspection.

The input space of a program is often so large that it is infeasible to iterate over every single input sequentially and verify for each input that the program behaves correctly. Because of this, fuzzers use various approaches to search through the input space. \emph{Grey-box fuzzing} uses observations during the execution of a program to guide the search for relevant inputs. For example, one way to do this is to use code coverage as a metric: if an input takes a not-seen-before execution path through the code, it is more likely to be an interesting input than if it had taken an execution path seen many times before.

The gray-box fuzzer afl \cite{afl} relies on observation of the execution path taken by the program to guide its generation of new inputs. It uses inputs that trigger an unseen execution path to generate new inputs, and discards others. In order to observe the execution of a program, code needs to be \emph{instrumented}. This is usually done at compile time, by automatically inserting instrumentation instructions in the program code. If source code is not available, afl can also instrument at runtime by running the program in a modified emulator that can insert instrumentation instructions on the fly.

\section{Memory sanitizers}
Memory safety violations \cite{sok-sanitizers} are programming errors where a pointer accesses a memory location outside the bounds of the intended target (a \emph{spatial} safety violation) \textit{or} when a pointer accesses memory at a point in time where that memory is not valid, for example because it has been deallocated (a \emph{temporal} safety violation).

Memory safety issues generally do not cause the program to crash directly. Instead, an out-of-bounds error on the stack just overwrites some other variables. It is not until those variables are eventually used that unintended behavior or a crash occurs. This "delay" between memory error and altered behavior can make it harder to efficiently search the input space. \emph{Sanitizers} can remove this delay by directly aborting the code execution at the moment a memory error occurs.

Sanitizers add instrumentation to software programs to check for various sorts of issues like memory safety violations. To do this, most sanitizers instrument code at compile time. There are also sanitizers that can instrument binaries at runtime, or when running code inside an emulator. A recent analysis by Song et al.\ \cite{sok-sanitizers} identifies two different kinds of sanitizers for memory safety violations:

\begin{enumerate}
    \item Location-based sanitizers, which detect memory accesses to invalid regions.  
    \item Identity-based sanitizers, which detect memory accesses that do not target their intended referent.
\end{enumerate}

For both approaches, code is predominantly instrumented at compile time. In the next two subsections we consider each of these approaches and provide an overview of the tools available when trying to sanitize a binary program.

\subsection{Location-based sanitizers}
\begin{figure}[!htb]
    \def\svgwidth{\textwidth}
    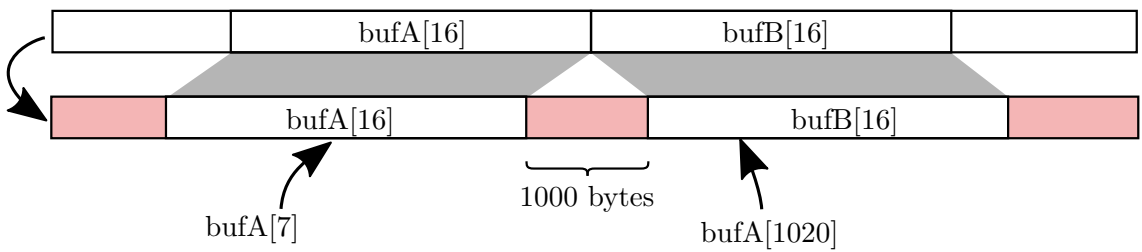
    \caption{Location-based sanitizers insert redzones in between buffers. When memory is accessed, the target memory location is checked. If it is a redzone, an error is raised.}
    \label{fig:redzones}
\end{figure}

Location-based sanitizers work by maintaining metadata on each byte of memory, and inserting large spaces of invalid memory between variables in order to detect overflows. This can for example be done by inserting large areas of memory that are marked as 'invalid', called \emph{redzones}, illustrated in figure \ref{fig:redzones}, or by placing each allocation on its own guard page. Examples of tools using this approach are AddressSanitizer \cite{asan} (ASan) and MemorySanitizer\cite{msan} (MSan), which are also often used in combination with afl.

Using redzones or guard pages requires large amounts of memory, which presents problems if the program is being run in an emulator. afl \cite{afl} uses a modified version of qemu \cite{afl-qemu} to collect execution traces of binaries that could not be instrumented with coverage instructions at compile time. qemu is not capable of emulating the required memory efficiently. Consequently, ASan, MSan, and other sanitizers are listed as incompatible with afl in qemu mode \cite{afl-asan-notes}.

The location-based sanitizer QAsan \cite{qasan} attempts to work around this issue. It modifies qemu to hook memory allocation functions like \texttt{malloc}, and then passes information by syscall or custom instruction to outside qemu, where a sanitizer runs. This way memory can be allocated outside the emulated environment, thus avoiding emulation issues with qemu.
 
RetroWrite \cite{retrowrite} takes a different approach to instrumenting binaries. Instead of running code in an emulator and instrumenting at runtime like QAsan does, it statically instruments a binary before execution by analyzing the assembly and inserting instrumentation at various points in the code. It is able to distinguish between pointers and integer values by (ab)using the relocations stored in the executable file. These relocations are used to adjust pointer offsets if the binary is loaded into a different region of memory than originally intended when compiling the program. Relocation data is not always available; for example, on Linux-based systems it is only stored in so called \emph{Position Independent Executables} (PIEs). PIEs are commonly used on relatively modern systems for Address Space Layout Randomization. Although RetroWrite performs analysis of the binary, it does not modify stack frames themselves and only inserts redzones in between frames. This means that it can catch an overflow if and only if it overflows far enough such that it attempts to read or write data in the redzone. To improve performance, redzones are only inserted if presence of a stack canary is detected. If no stack canary is detected, it is assumed that the compiler was able to statically prove that overflows in the function are impossible.

\subsubsection{Problems with location-based sanitizers}
Location-based sanitizers are not able to detect every overflow. Consider the example used in figure \ref{fig:redzones}. Location-based sanitizers would not alert on accessing \texttt{bufA[1020]}, because it accesses the valid memory location associated with \texttt{bufB[4]}. However, this is often not a problem as the chances of this happening are relatively small.

Inserting redzones when source code is not available is much more difficult for stack allocations compared to heap allocations. In contrast to heap-based allocations, which are allocated at runtime, variables on the stack are ordered and positioned at compile time such that their position is determined by an addition of the stack pointer and some constant offset. In order to be able to change this positioning, a location-based sanitizer for binaries would have to analyze the assembly code, re-construct the original layout of the stack frame, insert redzones into this frame, and then recompile the assembly with these new offsets. None of the tools mentioned in this section are able to sanitize stack variables at a level more granular than an entire stack frame if the source code is not available.

When source code is not available, global variables share the same issues as stack variables for location-based sanitizers. Additionally, global variables are not allocated on stack frames so there is no natural boundary between a group of variables. Therefore a location-based sanitizer would not be able to insert redzones between the frames without first distinguishing all global variables \emph{and} all accesses to these variables, and then rewriting the entire program to account for the redzones inserted between the global variables. There is currently no tool that attempts to do this.\label{locsan-globs}

\subsection{Identity-based sanitizers}\label{intro-id}
Whereas location-based sanitizers only check whether \textit{valid} memory is accessed (and therefore need to insert large amounts of invalid memory between variables to be able to detect memory safety violations), identity-based sanitizers check whether the \textit{correct} memory (i.e. the memory with the same "identity" as the pointer that is being used to access it) is being accessed. Because of this, identity-based sanitizers can detect more memory safety violations than location-based sanitizers. This additional precision of identity-based sanitizers comes at the cost of performance, as identity-based sanitizers are often slower than location-based sanitizers \cite{sok-sanitizers}.

\begin{figure}[!htb]
    \def\svgwidth{\textwidth}
    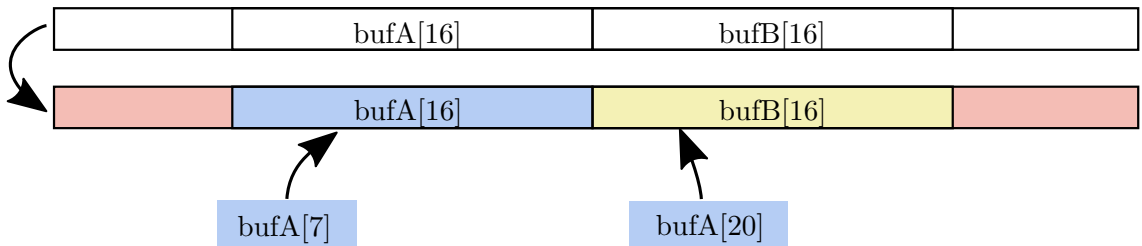
    \caption{Identity-based sanitizers keep track of the "identity" of memory regions and pointers, here indicated with different colors. When memory is accessed, the identity of the pointer is compared to the identity of the memory region. If they are not equal, an error is raised.}
    \label{fig:identity-colors}
\end{figure}

Identity-based sanitizers maintain metadata (the "identity") for each pointer in use by the program. Upon encountering a memory access, the sanitizer will check the metadata to see if the memory access is valid. Figure \ref{fig:identity-colors} shows such an approach where the metadata for a pointer is a color. An access check would just require checking the equality of the pointer color and the color of the memory region. \texttt{bufA[7]} would access a blue memory region using a blue pointer, so this is allowed. \texttt{bufA[20]} would access a yellow memory region using a blue pointer, which is not allowed. Note, however, that the color of the memory region does not necessarily need to be stored anywhere and could, for example, be derived from the address of the memory location instead.

A common choice for metadata is to store a tuple $(\textit{base}, \textit{size})$ where \textit{base} is the lowest address in the memory region and \textit{size} is the length of this memory region. This requires no metadata to be stored for the memory, as an access check for address \textit{addr} only requires verifying that $\textit{base} \leq \textit{addr} < \textit{addr} + \textit{size}$.

Early identity-based bounds checking, like Bcc \cite{bcc} or the subsequent Safe-C \cite{safe-c} implement bounds checking for C by redefining pointers to be a type that contains not only the actual pointer value, but also metadata for the pointer. In the case of Safe-C, this metadata contains:

\begin{enumerate}
    \item base address of the pointer
    \item size of the memory region pointed to by the pointer
    \item storage class (heap, stack, global -- for spatial checks)
    \item capability (forever, never -- for temporal checks)
\end{enumerate}

Since the memory representation of a pointer is changed, any library used by the program has to be recompiled to support these new pointer types. This presents interoperability issues with existing libraries. Because of these issues, much of the work after the tools described in the previous paragraph focuses on ways to improve this interoperability and abandons the idea of true fat pointers where metadata is stored inside the pointer. Some approaches move the metadata out of the pointer into a separate structure on the stack or heap \cite{pcheck-c}: giving each pointer a separate "guard" object containing metadata, and each pointer array a separate "guard" array of metadata keeps pointer representation the same. This enables interoperability with libraries that were not compiled to support new pointer representations.

Research more recent than the papers discussed above tries to eliminate the overhead of having to update the pointer metadata constantly. For example, Baggy Bounds Checking \cite{bbc}, like the independently developed PAriCheck \cite{paricheck}, allocates memory smartly to ensure that all metadata is encoded in the pointer itself, while also allowing the pointer to be dereferenced without requiring additional operations. Because of this encoding, it is compatible with libraries that are compiled without support for the sanitizer, as long as source code for the program itself is available. It does not sanitize stack- or global variable accesses.

The MAI (Memory Access Integrity) sanitizer \cite{mai} is the only identity-based memory sanitizer that works on binaries, and attempts to recover a fine-grained memory structure from the access patterns observed during execution. It is implemented on top of an abstraction framework which translates x86 instructions into an intermediate representation that can more easily be consumed by a sanitizer.

MAI builds a tree-like representation of memory, where a tree node is associated with a memory region, and each of its children represent a smaller part of that memory region. These regions are determined by looking at the "first write" of that particular region. This relies on the assumption that memory is always initialized, and when it is initialized this is done field-by-field. It is unclear how well this interacts with compiler optimizations like inlined calls to \texttt{memset} or \texttt{memcpy}. For global variables (which can be initialized at compile time instead of runtime), MAI can use manually entered data. Because of its precision, MAI also has trouble with \emph{union fields} (fields that may be accessed as multiple different types, potentially of different sizes). If the first access does not access the largest type in the union, the inferred size of the memory region will be too small.

MAI is also unable to handle C strings (i.e. strings terminated by a NULL byte) in some cases, as the zero-byte written at the end might be interpreted as a separate 1-byte field. Thus, when the program reads through the entire string, it fails when reading the NULL byte.
\chapter{The IdSan sanitizer}\label{design}
While location-based sanitizers require relatively little computation besides validating every memory access and marking memory regions upon allocation and deallocation, identity-based sanitizers need to do more tracking in between. Pointer metadata needs to be tracked as values move around registers and are written to and read from memory. In this chapter we describe the metadata that we are storing (section \ref{pointer-metadata}), the information that the sanitizer is tracking (section \ref{sanitizer-state}) and the operations on that information (section \ref{pointer-ops}), and finally the reasons for this specific design (section \ref{design-considerations}).

\section{Overview}
IdSan maintains a state, which contains metadata on all pointers IdSan is tracking. It consumes \emph{operations} and produces either a confirmation that the operation was successful or a memory safety violation error. The operations are an abstraction over raw assembly instructions. Because we are not directly operating on assembly instructions, we can independently map assembly instructions to sanitizer operations in a separate component.

\subsection{Pointer metadata}\label{pointer-metadata}
Since we are implementing bounds checking for binaries and not for source code, we do not have interoperability issues usually associated with fat pointers. These interoperability issues stem from the fact that precompiled binaries might be linked in with the program, and instrumentation is missing from those binaries. Since we are not dealing with source code at all, our approach can use the same general approach as the sanitizer by Patil and Fischer \cite{pcheck-c}, which stored metadata in separate objects.

\begin{definition}
    The identity of a pointer is defined as a tuple $(\text{base}, \text{length})$, where base refers to the lowest address in memory that this pointer may access, and length the number of bytes following that address that can also be accessed. Any memory address between base and base+length can be accessed by the pointer.
\end{definition}

\begin{example}
    For a pointer to a 10-character \texttt{char} array, the identity might look like $(\text{0xFF80, 10})$ indicating that this pointer points to a memory region between 0xFF80 and 0xFF8A (excluding 0xFF8A).
\end{example}

\begin{definition}
    We define "word" to mean a number that is as many bits wide as a pointer. For example, an x86-64 "word" would be 64 bits long\footnote{under Windows terminology this is usually called a QWORD}. While the exact size of a word is not relevant for the definition of our semantics, we will assume it to be 64 bits in the examples we give in this section.
\end{definition}

However, for reasons that will be explained in section \ref{memcpy-issues} we cannot associate a single identity directly with a single word. Instead, we need to consider a word to be a fixed number of bits, all of which might have different identities. Therefore, we have to define metadata for a word (or any $n$-bit value if we want to support registers of sizes other than a single word) more broadly.

\begin{definition}
    \label{value-metadata}
    The metadata for an $N$-bit value is defined as an ordered list of triples of the form $(\textit{start}..\textit{end}, \textit{identity})$. In other words, we are storing which pointer each set of bits in a value belongs to ($\textit{identity}$), as well as which bits from the pointer they represent ($\textit{start}..\textit{end}$).

    If there exists a range which does not contain bits from a pointer, this is represented as $(0..\textit{length}, -)$. The total number of bits stored in the list should always be equal to $N$, i.e. the sum of $\textit{end} - \textit{start}$ for all triples should be equal to $N$.
\end{definition}

In the case of a register, the associated metadata is simply a single metadata object of the same size as the register. Usually this will be the size of a word, however in the case of vector registers the metadata object might be much larger. For example, for an ARM NEON register \cite{neon} the metadata would be 128 bits wide.

We consider memory to be a large array of words, and consequently store a single metadata object per word of memory. This is efficient in the common case where pointers are aligned: a memory read or write of a pointer can be done by simply copying the metadata of the corresponding word in memory. When pointers are unaligned, metadata of multiple memory words might need to be sliced and concatenated to produce the metadata for the value that is being read.

\begin{example}
    Consider a 64-bit pointer $A$, stored fully in a register $R_0$. The metadata for register $R_0$ is then represented as: $[(0..64, A)]$.
\end{example}

\begin{example}
    Consider a misaligned\footnote{This alignment can, of course, be perfectly valid in the context of the architecture this code is running on. However for our purposes we can consider it misaligned in the sense that only part of the pointer is loaded into a register.} 64-bit pointer $A$, which during a \texttt{memcpy} operation only has its lower 32 bits stored in a register $R_0$. The metadata for register $R_0$ is then represented as: $[(0..32, A), (0..32, -)]$.
\end{example}

\begin{example}
    Consider three 64-bit pointers $A, B, C$, where the last 48 bits of $A$, all of $B$ and the first 16 bits of $C$ are stored in a single 128-bit vector register $V_0$. The metadata for register $V_0$ is then represented as: $[(16..48, A), (0..64, B), (0..16, C)]$.
\end{example}

\subsection{State}\label{sanitizer-state}
\begin{definition}
    The sanitizer state is a triple $(\textit{regs}, \textit{memory}, \textit{stack})$. 
    
    \begin{enumerate}
        \item \textit{regs} contains metadata for each register
        \item \textit{memory} is a finite function $\mathbb{N} \rightarrow \textit{metadata}$ mapping each word in memory to its metadata. For memory that does not have any metadata, the output is a new empty metadata object of one word
        \item \textit{stack} is a shadow stack which contains the following information for each stack frame:
        \begin{enumerate}
            \item The memory address of the frame
            \item The size of the frame
            \item Whether the frame belongs to a \texttt{malloc}-like function and if it does also the number of bytes that were requested to be allocated
            \item The program counter when this frame was pushed onto the stack
        \end{enumerate}
    \end{enumerate}
\end{definition}

The shadow stack is used for multiple purposes:
\begin{enumerate}
    \item To be able to make educated guesses on the maximum size of stack variables, if that information is missing from the manually provided information
    \item To provide stack traces when detecting a memory error
    \item To be able to correctly set the metadata for the stack pointer, which should encompass the 2 topmost stack frames because of stack-allocated function arguments as described in section \ref{stack-args}.
\end{enumerate}

\subsection{Operations}\label{pointer-ops}
\begin{definition}
    We define the following operations for pointers:
    \begin{enumerate}
        \item \textbf{Copy}(\textit{src}, \textit{dst}, \textit{transform}): This operation simply copies metadata from one register to another. Optionally, this metadata can be transformed by one of the following operations:
        \begin{enumerate}
            \item \textbf{Nothing}
            \item \textbf{ShiftLeft}(\textit{n})
            \item \textbf{ShiftRight}(\textit{n})
            \item \textbf{RotateLeft}(\textit{n})
            \item \textbf{RotateRight}(\textit{n})
        \end{enumerate}
        \item \textbf{Clear}(\textit{r}): This operation clears all metadata for register \textit{r}
        \item \textbf{BitFieldCopy}(\textit{dst}, \textit{src}, \textit{src\_off}, \textit{size}, \textit{dst\_off}, \textit{clear}): Copies \textit{size} bits starting at \textit{src\_off} from register \textit{src} to register \textit{dst} at bit offset \textit{dst\_off}. If \textit{clear} is set, all other bits are set to 0. If unset, all other bits in \textit{dst} are unchanged.
        \item \textbf{Combine}(\textit{a}, \textit{b}, \textit{dst}): This operation "combines" metadata from two registers \textit{a} and \textit{b} into a single result, and stores this in the \textit{dst} register.
        \item \textbf{Access}(\textit{op}, \textit{addr}, \textit{via}, \textit{reg}, \textit{n}): Reads or writes (depending on \textit{op}) \textit{n} bytes from memory at \textit{addr}, using the pointer in \textit{via}, into or from register \textit{reg}. The condition $\text{addr} \geq \text{base} \wedge \text{addr} + \text{size} \leq \text{base} + \text{length}$ must hold for the operation to be valid.
        \item \textbf{CreateStackPointer}(\textit{pc}, \textit{sp\_off}, \textit{base}, \textit{dst}): Store a new pointer to a variable on the stack at offset \textit{sp\_off}, starting at \textit{base}, with the program counter at \textit{pc}, into register \textit{dst}.
        \item \textbf{CreateGlobalPointer}(\textit{pc}, \textit{offset}, \textit{dst}): Store a new pointer to a global variable at offset \textit{offset}, with the program counter at \textit{pc}, into register \textit{dst}.
        \item \textbf{StackChange}(\textit{offset}, \textit{new\_base}): Indicates that the stack pointer was modified to a new value. Usually this only happens at the start and end of a function, but it can also occur when variable-length arrays are allocated on the stack.
        \item \textbf{FunctionCall}(\textit{sp}, \textit{target\_pc}, \textit{arg0}): Indicates that a function call was made.
        \item \textbf{Return}(\textit{r}, \textit{val}): Indicates that a return instruction was encountered and that a function call is returning \textit{val} in register \textit{r}.
    \end{enumerate}
\end{definition}

\section{Design considerations} \label{design-considerations}

\subsection{Memory copying}\label{memcpy-issues}
At first glance, one might assume that storing metadata similar to the $(\textit{base}, \textit{size})$ fat-pointer implementations mentioned in section \ref{intro-id} might be sufficiently precise. However, this model doesn't work well on modern instruction set architectures. Operations like \texttt{memcpy} are unaware of the contents of the memory that they are copying. This means that the implementation does not adhere to alignments or structure that was specified in the original code. 

\begin{figure}[!htb]
    \begin{center}
    \def\svgwidth{0.5\textwidth}
    %% Creator: Inkscape inkscape 0.92.5, www.inkscape.org
%% PDF/EPS/PS + LaTeX output extension by Johan Engelen, 2010
%% Accompanies image file 'memcpy.pdf' (pdf, eps, ps)
%%
%% To include the image in your LaTeX document, write
%%   \input{<filename>.pdf_tex}
%%  instead of
%%   \includegraphics{<filename>.pdf}
%% To scale the image, write
%%   \def\svgwidth{<desired width>}
%%   \input{<filename>.pdf_tex}
%%  instead of
%%   \includegraphics[width=<desired width>]{<filename>.pdf}
%%
%% Images with a different path to the parent latex file can
%% be accessed with the `import' package (which may need to be
%% installed) using
%%   \usepackage{import}
%% in the preamble, and then including the image with
%%   \import{<path to file>}{<filename>.pdf_tex}
%% Alternatively, one can specify
%%   \graphicspath{{<path to file>/}}
%% 
%% For more information, please see info/svg-inkscape on CTAN:
%%   http://tug.ctan.org/tex-archive/info/svg-inkscape
%%
\begingroup%
  \makeatletter%
  \providecommand\color[2][]{%
    \errmessage{(Inkscape) Color is used for the text in Inkscape, but the package 'color.sty' is not loaded}%
    \renewcommand\color[2][]{}%
  }%
  \providecommand\transparent[1]{%
    \errmessage{(Inkscape) Transparency is used (non-zero) for the text in Inkscape, but the package 'transparent.sty' is not loaded}%
    \renewcommand\transparent[1]{}%
  }%
  \providecommand\rotatebox[2]{#2}%
  \newcommand*\fsize{\dimexpr\f@size pt\relax}%
  \newcommand*\lineheight[1]{\fontsize{\fsize}{#1\fsize}\selectfont}%
  \ifx\svgwidth\undefined%
    \setlength{\unitlength}{470.25621634bp}%
    \ifx\svgscale\undefined%
      \relax%
    \else%
      \setlength{\unitlength}{\unitlength * \real{\svgscale}}%
    \fi%
  \else%
    \setlength{\unitlength}{\svgwidth}%
  \fi%
  \global\let\svgwidth\undefined%
  \global\let\svgscale\undefined%
  \makeatother%
  \begin{picture}(1,1.00353145)%
    \lineheight{1}%
    \setlength\tabcolsep{0pt}%
    \put(0,0){\includegraphics[width=\unitlength,page=1]{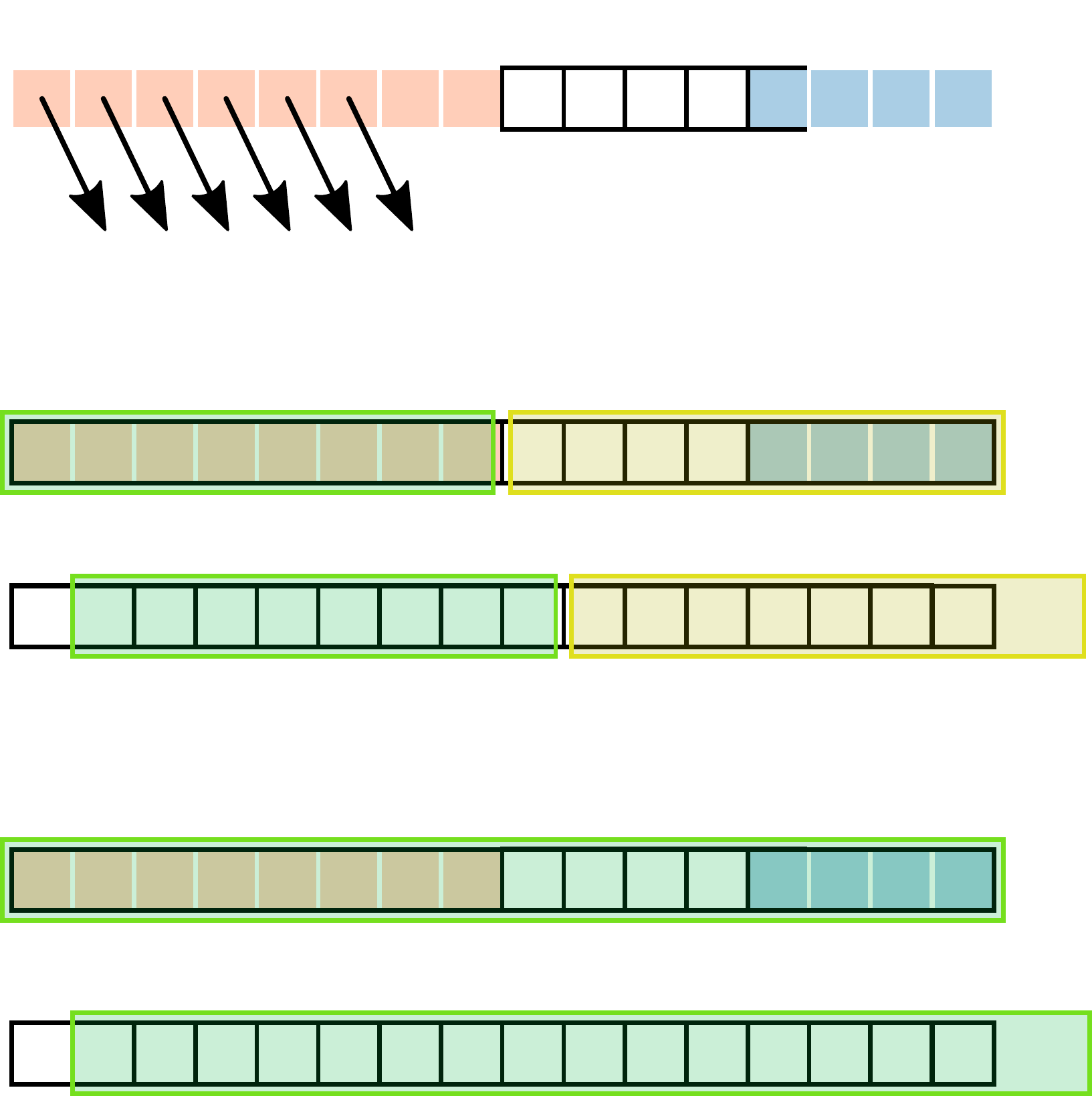}}%
    \put(0.00624283,0.25186885){\color[rgb]{0,0,0}\makebox(0,0)[lt]{\lineheight{1.25}\smash{\begin{tabular}[t]{l}\textbf{3. memcpy using vector registers}\end{tabular}}}}%
    \put(0.00598719,0.64350537){\color[rgb]{0,0,0}\makebox(0,0)[lt]{\lineheight{1.25}\smash{\begin{tabular}[t]{l}\textbf{2. memcpy using registers}\end{tabular}}}}%
    \put(0.00529524,0.95855597){\color[rgb]{0,0,0}\makebox(0,0)[lt]{\lineheight{1.25}\smash{\begin{tabular}[t]{l}\textbf{1. byte-for-byte memcpy}\end{tabular}}}}%
    \put(0,0){\includegraphics[width=\unitlength,page=2]{memcpy.pdf}}%
  \end{picture}%
\endgroup%

    \end{center}
    \caption{Three \texttt{memcpy} implementations. Each square represents a single byte of memory. The red and blue colored squares are (parts) of two different pointers. The red pointer is aligned, while the blue pointer is 4 bytes off alignment.}
    \label{fig:memcpy}
\end{figure}

Modern compilers are able to detect \texttt{memcpy}-like loops and will actively replace those with calls to \texttt{memcpy}. For example, a \texttt{for}-loop that copies 64 shorts to another array might be unrolled and optimized into 8 128-bit vector moves. Therefore, we need to be able to deal with pointers that end up being copied by (an inlined) \texttt{memcpy} to different locations in memory, as we do not have the ability to instrument this at compile time. As far as implementations for \texttt{memcpy} are concerned for our use, there are 3 cases we can distinguish as illustrated in figure \ref{fig:memcpy}:

\begin{enumerate}
    \item Byte-for-byte copying means that a single pointer is moved piece-by-piece over multiple operations. This implementation still allows us to do tracking as described above, since a register will never contain more than a single pointer. However, this implementation is rarely used as it is the most inefficient way to copy memory.
    \item Using the full register size, multiple bytes can be copied in a single operation. This is much more efficient, however for the tracking described above it requires that all pointers are aligned to the number of bytes that we are copying at a time, such that we never copy two partial pointers at the same time. In figure \ref{fig:memcpy} this issue is illustrated by the blue area indicating a pointer, which is not aligned such that the full pointer will be copied in a single operation.
    \item Vector registers are large registers intended for performing "single instruction, multiple data" (SIMD) operations, like AVX and SSE \cite{x86-simd} on x86-64 or NEON on ARM \cite{neon}. These registers are much bigger than the pointer size of the architecture. For example, NEON registers on ARM are 128 bits wide, which means it is possible to copy up to 3 (partial) 64-bit pointers in a single operation.
\end{enumerate}

This problem is unique to identity-based sanitizers. Location-based sanitizers only ensure that the memory that is accessed is valid, for which accurate tracking of pointers is not required and therefore memcpy operations would not need to be instrumented beyond verifying that the source and destination memory is valid.

Memory copying operations are commonly used in almost any kind of program. It is likely that at some point a memory structure containing a pointer will be copied by \texttt{memcpy} or a \texttt{memcpy}-like operation, so we need to use a more fine-grained approach where a single metadata object can hold information on multiple pointers.

To solve this problem, we keep separate metadata for each bit in a value as described in Definition \ref{value-metadata}. By doing this, we can handle multiple pointers in a single value as well as a partial pointer in a value.

% MAI TODO: {Does MAI account for this at all? -- once again need source to confirm as this is missing in paper.}

\subsection{Stack pointer identification}\label{stack-var-identification}
Generally, instruction sets do not have separate instructions for "taking a pointer" to an element on the stack. One would need access to the source code to conclusively determine this. Many location-based sanitizers are either able to mitigate this limitation by instrumenting at compile time and inspecting the source code, or simply ignore stack operations entirely.

We use a heuristic to detect pointers to the stack, where we consider any operation of the form \texttt{add $\texttt{r}_{\texttt{dst}}$, sp, N} (where N is some constant number) to be the computation of a pointer to a value on the stack. While compilers are not required to use this operation, and might use other tricks, it is likely that it is used for two reasons:

\begin{enumerate}
    \item If a pointer needs to be passed to another function, using \textit{sp}+\textit{N} is the most efficient way to compute it
    \item While it might be possible to iterate over an array by computing the address of an element as $\textit{sp}+\textit{N}+\textit{index}$ directly, it is more efficient to pre-compute some $\textit{p}=\textit{sp}+\textit{N}$ and then use $\textit{p}+\textit{index}$ to access individual elements of the array. While it is also possible to start iteration at $\textit{index'}=\textit{N}$ instead of 0 and then simply use $\textit{sp}+\textit{index'}$ to access an array element, this is also unlikely to happen since all bounds checks would also need to be altered. This does of course not prevent the compiler from pre-computing $\textit{N}+\textit{index}$ at compile time, but in that case there would be no way to go out of bounds either so it can be safely ignored.
\end{enumerate}

One exception to this is when the offset is too big to store in the instruction itself. For example, in the AArch64 instruction set an instruction of the form \texttt{ADD sp, imm} can only contain a 12-bit immediate value, which has a maximum value of 4095. For any offset above that, the compiler has to generate \textit{two} addition instructions. To accommodate for this, stack pointer identification can be configured on a function-by-function basis to be delayed until the first memory access instead of the first occurrence. The delayed identification is also enabled automatically when the immediate value 4095 is added to the stack pointer, which allows IdSan to handle most cases automatically.

\subsection{Global pointer identification}
By default (for the AArch64 targets) \texttt{gcc} and \texttt{clang} load global pointers by first invoking the \texttt{adrp} instruction to load the base address of a page in a register, then adding an immediate value to compute the actual pointer. This makes identification of global pointers much easier and more reliable. Other instruction sets might not have such an instruction, which makes identification more difficult. In such cases, a solution might be to rely on relocation entries like RetroWrite \cite{retrowrite} or the \texttt{.got} section.

\subsection{Stack-allocated function arguments}\label{stack-args}
Functions are not limited to accessing only their own stack via the stack pointer. For example, on an AArch64 architecture the first 8 arguments to a function are stored in registers r0-r7. A ninth argument will be stored in the stack frame of the caller. The callee will then access this argument directly by using the offset of the argument relative to the stack pointer, thus accessing something outside its own stack frame.

Enforcing that functions only access their own stack frame would produce false positives, so we have no choice but to allow any function to access the two topmost stack frames (including their own stack frame).

\subsection{Swapped memory addressing}
Usually, the AArch64 load and store instructions take arguments of the form \texttt{LDR target, [ptr, offset]} where \texttt{ptr} is the register containing a pointer and \texttt{offset} is the register containing the offset (or index) that is being accessed. However, in some cases the compiler will swap these registers and instead put the register containing the offset in \texttt{ptr} and the register containing the pointer in \texttt{offset}. Because of this, we actually have to check the metadata for two registers before we can generate the \textbf{Access} operation. If our metadata indicates that \texttt{ptr} is in fact not a pointer, we will attempt to use \texttt{offset} as \textit{via} in the \textbf{Access} operations instead.

\subsection{Invalid memory accesses}\label{invalid-mem-access}
It can be advantageous to perform incorrect memory accesses to improve performance. For example, in the C standard library the \texttt{strlen} function might read memory in aligned chunks of 8 bytes, no matter where exactly the string ends. It inspects all 8 bytes at the same time, and directly skips to the next 8 bytes if no NULL byte is found. Since memory pages are aligned to at least 8 bytes, this will never result in a crash.

To mitigate this issue, a relaxation of the sanitizer rules that allows aligned reads of up to 7 bytes past the end of an object can be enabled on a function-per-function basis. These issues are optimizations that a developer has to specifically write code for, so we suspect that these cases will be extremely rare outside the standard library.

\subsection{Missing information}
The sanitizer needs to be initialized with a listing of all memory locations where pointers are stored, and of all variables any function stores on the stack. For missing information of pointers to global variables, the sanitizer simply treats the memory as a pointer to nothing (i.e. $[(0..64, -)]$). If information about variables on the stack is missing, the sanitizer will infer the pointer to simply point to the entire stack frame. This way, some rough out-of-bounds checking can still be done even if the program is not fully annotated.

There are multiple ways to handle missing information. The easiest way is to simply abort the program as soon as we encounter a pointer operation where we have no metadata on the pointer. However, this requires us to painstakingly annotate every single variable in every single function of the program being sanitized. While exporting these annotations from a reverse engineering tool like IDA \cite{ida} or radare2 \cite{radare2} could help, it still requires manual verification of the output.

Another way is to simply ignore operations involving missing metadata. However, this would negate most of the annotation work we are doing, as a single misbehaving pointer that we are not able to track could still overwrite anything it wants.

Our approach aborts when we encounter a pointer operation where we have no metadata on the pointer if and only if it accesses a region where we are tracking other pointers. This combines the best of both worlds: memory access errors are reported, but the program can be annotated incrementally such that you can start with the most interesting code and selectively add other regions of interest.

\section{Example}
Consider the following example program, which moves a few values between registers or memory:

\begin{lstlisting}[language=c]
    mov r0, r1
    mov [r0+0x08], r1
    mov [r0+0x10], r1
\end{lstlisting}

We assume that the register $r_3$ contains a pointer $[(0..64, 0xff10..0xff20)]$, and the value of $r_3$ is $0xff10$.

\begin{enumerate}
    \item Initially our state is $((r_0: [(0..64, -)], r_1: [(0..64, \textit{0xff10}..\textit{0xff20})]), [], -)$
    \item We translate \texttt{mov r0, r3} to \textbf{Copy}($r_0$, $r_3$, \textbf{Nothing}) and apply it to our state
    \item \textbf{Copy} just copies metadata to another register, so our new state is \\ $((r_0: [(0..64, \textit{0xff10}..\textit{0xff20})], r_1: [(0..64, \textit{0xff10}..\textit{0xff20})]), [], -)$
    \item We then encounter \texttt{mov [r0+0x08], r1}, which we translate as \textbf{Access}(\textbf{WriteTo}, 0xff18, $r_0$, $r_1$, 8)
    \item Now, we need to verify that $\text{addr} \geq \text{base} \wedge \text{addr} + \text{size} \leq \text{base} + \text{length}$ holds. By substitution, we can see that $0xff18 \geq \textit{0xff10}$ and $0xff18 + 8 = \textit{0xff20} \leq \textit{0xff20}$
    \item Finally, we update our state to $(..., [0xff18: [(0..64, \textit{0xff10}..\textit{0xff20})]], -)$ to track the pointer that we just stored
    \item We then encounter \texttt{mov [r0+0x10], r1}, which we translate as \textbf{Access}(\textbf{WriteTo}, \textit{0xff20}, $r_0$, $r_1$, 8)
    \item However this time $\textit{0xff20} + 8 = 0xff28 \nleq \textit{0xff20}$. This means that a memory safety violation has occurred, so abort execution
\end{enumerate}
\chapter{Implementation}\label{implementation}
Our implementation is focused on providing an easily modifiable and extensible tool. Therefore, we have not opted for a tight integration into for example unicorn, which could have improved performance but would have made supporting \texttt{ptrace} or a different instruction set impossible.

Our tool sits between afl and the binary in question as a testing harness. In unicornafl mode, it runs the full unicorn emulator inside the harness and manually maps an ELF binary into the emulator memory space. In \texttt{ptrace} mode, the harness launches the program as a separate process and attaches to it via the Linux \texttt{ptrace} syscall.

IdSan can also run in 'standalone' mode, so it is possible to use any other black-box fuzzer without any modifications to IdSan. Switching to another gray-box fuzzer is also possible, however this would likely require switching from the unicornafl emulator to unicorn itself, and implementing the instrumentation that the fuzzer requires.

\begin{figure}[!htb]
    \def\svgwidth{\textwidth}
    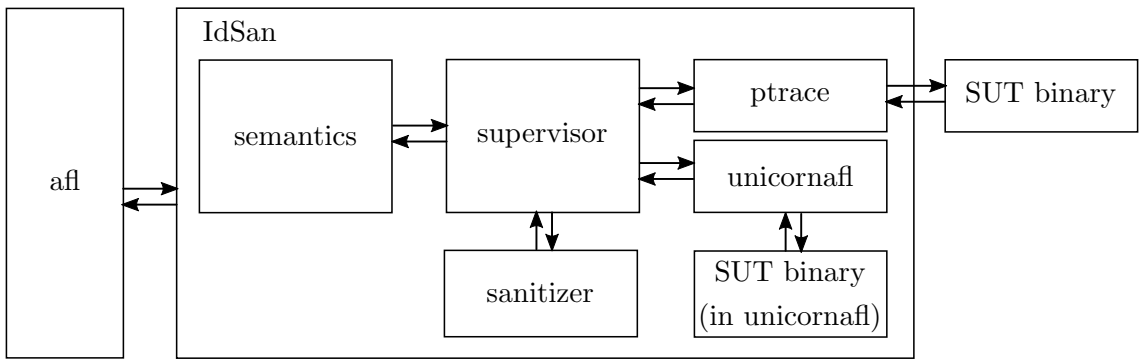
    \caption{The architecture of our implementation}
    \label{fig:code-arch}
\end{figure}

As demonstrated in figure \ref{fig:code-arch}, IdSan is split into 5 parts. It can work via either unicornafl or the Linux syscall \texttt{ptrace}:

\begin{enumerate}
    \item The supervisor handles all communication between the components. 
    \item The 'semantics' component (\texttt{src/arch/arm64.rs}) receives information on the processor state and the instruction that is about to execute, and translates this into operations that the sanitizer understands.
    \item The 'sanitizer' component (\texttt{src/sanitizer/*.rs}) is responsible for keeping the tracking metadata from section \ref{pointer-metadata} up to date. It stores information on all registers and memory. It receives operations to execute from the supervisor.
    \item The ptrace (\texttt{src/supervisor/ptrace.rs}) and unicornafl (\texttt{src/supervisor/unicorn.rs}) components contain backend-specific code to instrument binaries running via ptrace or in the unicorn emulator.
\end{enumerate}

\section{Limitations}
The length of the ordered list described in section \ref{pointer-metadata} turns out to be a performance issue, as this list can quite quickly grow in size. Therefore, in the actual implementation we limit the length of the ordered list to 6. If the length becomes longer than 6, we consider the register to no longer be storing any pointer. This keeps the amount of memory required for the metadata more manageable.

\section{AArch64}
IdSan implements semantics for the AArch64 (ARMv8) architecture. We chose this architecture because of its relatively nice instruction set compared to x86-64, as well as a slight interest from Riscure for this architecture. We will provide an overview of how AArch64 instructions are mapped onto the pointer operations defined in section \ref{pointer-ops}.

As we are only interested in tracking pointers as they move through the program, the most common reason for ignoring an instruction or clearing the metadata of the target register of an instruction is that it is an operation that is unlikely to ever be executed on a pointer or be used when copying memory that might contain a pointer from one location to another. 

\begin{enumerate}
    \item \texttt{MOV}, \texttt{MVN}, \texttt{MOVK}, \texttt{MOVZ}, \texttt{MOVN}, \texttt{CINV}, \texttt{CNEG}, \texttt{NEGS}, \texttt{NGCS} and \texttt{NGS} are mapped to \textbf{Copy} operations if the source operand is a register, or \textbf{Clear} if the source operand is an immediate value, as immediate values are only 12 bits and cannot be used as pointers.
    \item \texttt{ADRP} is mapped to \textbf{CreateGlobalPointer}, as it is always used by \texttt{gcc} (and likely other compilers as well) to compute the base of a global pointer.
    \item \texttt{ADD} and \texttt{SUB} mapped as follows:
    \begin{enumerate}
        \item \textbf{StackChange} if the instruction modifies the stack pointer
        \item \textbf{CreateStackPointer} if the instruction is an \texttt{ADD} instruction with one of the operands being the stack pointer, as explained in Section \ref{stack-var-identification}
        \item \textbf{Combine} if both operands are registers
        \item \textbf{Copy} otherwise, as the instruction is possibly an indexing operation with a constant offset \label{why-copy}
    \end{enumerate}
    \item \texttt{ADC}, \texttt{SBC}, \texttt{BIC}, \texttt{ORR}, \texttt{ORN}, \texttt{EOR}, \texttt{EON} are mapped to \textbf{Combine} if both operands are registers, \textbf{Copy} otherwise for the same reasons as in \ref{why-copy}
    \item \texttt{SMADDL} and \texttt{MADD} are multiply-add instructions that compute something of the form $(a * b) + c$, this could be used to compute the memory offset of a big structure in an array, i.e. it could be used to compute $\&\textit{array}[\textit{index}] = (\textit{index} * \textit{sizeof(array\_element)}) + \&\textit{array}$, so these are mapped to \textbf{Copy}
    \item \texttt{AND} is mapped to \textbf{Clear} if both operands are registers, \textbf{Copy} otherwise as an AND operation is sometimes used to zero out the lower bits of a pointer in order to align it to some multiple of 2
    \item \texttt{LSL}, \texttt{LSR}, \texttt{ROR}, \texttt{ASR}, \texttt{SHL} are mapped to \textbf{Copy}, with the appropriate shift applied as transformation
    \item \texttt{NEG}, \texttt{NOT}, \texttt{MRS}, various multiplication and division instructions, \texttt{CSET}, \texttt{CSETM} and \texttt{CCMN} are mapped to \textbf{Clear}, as it would not make sense to perform one of these operations on a pointer. For example, it does not make sense to write \texttt{4 * (char*)str} as the resulting pointer will not be pointing to anything useful
    \item \texttt{STP}, \texttt{LDP}, \texttt{LDPSW}, \texttt{LDXP}, \texttt{LDNP}, \texttt{STNP}, \texttt{LDAXP}, \texttt{STR[\{B,H\}]}, \texttt{LD[T]R[\{SW,H,SH,B,SB\}]}, \texttt{LDUR[\{B,H,SB,SH,SW\}]}, \texttt{LDXR[\{B, H\}]}, \texttt{ST\{T,L,U,A\}R[\{B,H\}]}, \texttt{LDAXRB}, \texttt{LDAXRH}, \texttt{LDAXR} as well as the exclusive store variants \texttt{ST[L]XP}, \texttt{STLXR[\{B,H\}]} and \texttt{STXR[\{B,H\}]}, are mapped to the appropriate (sequence) of \textbf{Access} operations
    \item The conditional operations \texttt{CSEL}, \texttt{CSINC}, \texttt{CSINV} and \texttt{CSNEG} are mapped to \textbf{Copy}, with the source set to the correct register depending on the condition code that is being checked
    \item \texttt{SBFIZ}, \texttt{SBFM}, \texttt{SBFX}, \texttt{SXTB}, \texttt{SXTH} and \texttt{SXTW} (sign-extended bitfield moves) are all mapped to \textbf{Clear}, because sign-extension on a pointer doesn't make much sense
    \item \texttt{BFM}, \texttt{BFI}, \texttt{BFXIL}, \texttt{UXTB}, \texttt{UXTH}, \texttt{UXTW}, \texttt{UBFM}, \texttt{UBFIZ} and \texttt{UBFX} (bitfield moves) are mapped to \textbf{BitFieldCopy} operations
    \item \texttt{BR} and \texttt{BLR} are mapped to \textbf{FunctionCall}
    \item \texttt{RET} is mapped to \textbf{Return}

    \item Many vector-specific operations are ignored (note that basic loads and stores to the vector registers $Q_n$ are implemented under \texttt{LDR}, \texttt{STR} and related instructions). Some instructions that were considered for implementation, but did not turn out to be necessary are:
    \begin{enumerate}
        \item the \texttt{LDn} and \texttt{STn} sets instructions as well as the \texttt{LDnR} set of instructions offer more fine-grained control over loading values into registers, but they don't seem to be used in memory copying operations
        \item \texttt{DUP}, \texttt{SMOV}, \texttt{EXTR} and \texttt{SRI} also don't appear to be commonly used in memory copying operations
    \end{enumerate}
    \item All floating-point operations are ignored as it does not make sense to store a pointer as a floating-point value, nor does it make sense to use floating point instructions to copy memory
    \item All comparisons, branches, and SHA, AES and CRC instructions as well as various system or hardware operations like \texttt{SVC}, \texttt{WFI} or \texttt{MSR} are ignored, as these do not interact with pointers or memory copying
    \item Some other instructions that were considered, but ultimately not implemented are: 
    \begin{enumerate}
        \item \texttt{REV}, which reverses the bytes in a register. A scenario where this instruction would be used on a pointer or during a memory copying operation seems unlikely, as long as no weird endianness changes are being done
        \item \texttt{ADR}, which computes an offset relative to the program counter. In theory this could be used to compute addresses to variables stored close (or in between) code, but compilers don't use this instruction for that purpose by default
    \end{enumerate}
\end{enumerate}
\chapter{Evaluation}\label{evaluation}
\section{Toy programs}
We wrote around 15 toy programs containing various memory safety issues, and made sure that IdSan worked correctly on each of the programs. These are very small programs that are intended to illustrate a single thing that the sanitizer might encounter and should alert on. For example, the toy program below uses a global variable. Depending on the length of the input, the \texttt{copy} function might overflow \texttt{target\_buffer} which should be detected by IdSan.

\begin{lstlisting}
unsigned char target_buffer[10];

void copy(unsigned char* from, unsigned long long length, unsigned char* to) 
{
    for(unsigned long long int i = 0; i < length; i++) 
    {
        to[i] = from[i];
    }
}

__attribute__((section(".text.startup"))) int entry_point(unsigned long long input_length, unsigned char* input) 
{
    copy(input, input_length, &target_buffer[0]);
    return 0;
}
\end{lstlisting}

We tested for the following scenarios:

\begin{enumerate}
    \item Global variable overflow
    \item Stack variable overflow
    \item Handling of \texttt{va\_args} (which requires the compiler to do strange stack pointer arithmetic)
    \item Handling of huge stack buffers (which makes single-instruction base-pointer addressing of the form \texttt{add sp, imm} impossible and instead requires the compiler to generate a sequence of additions)
    \item Functions with more than 8 arguments
    \item Copying pointers stored in memory using:
    \begin{enumerate}
        \item copying that uses bitshifts to combine 16-bit words together into 32-bit words
        \item byte-for-byte copying
        \item copying of pointers that are not aligned to 8-byte boundaries
    \end{enumerate}
    \item Array indexing via various operations like multiplications, array accesses, bitwise operations or additions
    \item An overflow that depends on a conditional \texttt{CSEL} instruction
    \item Memory copying via musl's \texttt{memcpy}
    \item Accessing \texttt{*(dest + ((end - start) - 10))}, where \texttt{end} and \texttt{start} are pointers (which requires the sanitizer to understand that subtracting two pointers gives an offset, and adding that offset to a pointer results in a valid pointer again)
\end{enumerate}

We believe these programs cover most code constructs that will be commonly encountered by IdSan.

\section{Real-world libraries}
We tested IdSan on a number of real-world libraries:

\begin{enumerate}
    \item \emph{LodePNG}, a single-file PNG encoder and decoder \footnote{https://lodev.org/lodepng/}
    \item \emph{yxml}, a single-file XML parser \footnote{https://dev.yorhel.nl/yxml}
    \item \emph{jsmn}, a single-file JSON parser \footnote{https://github.com/zserge/jsmn}
    \item The mp3-decoding functionality of \textit{LAME}\footnote{https://lame.sourceforge.io/}
\end{enumerate}

We verified that IdSan worked correctly on these libraries, and also fuzzed each library for a few hours. We did not discover any new vulnerabilities in the libraries. Unfortunately, because of the overhead of the sanitizer we were not able to fuzz LAME effectively, as we were only able to run around 1 execution per second.

To verify that IdSan is able to identify memory safety issues in bigger libraries as well, we introduced a vulnerability into LodePNG by replacing a bounds check:

\begin{lstlisting}
    // Before:
    if((size_t)((chunk - in) + chunkLength + 12) > insize || (chunk + chunkLength + 12) < in) {

    // After:
    if((size_t)((chunk - in) + chunkLength + 10) > insize || (chunk + chunkLength + 10) < in) {
\end{lstlisting}

\texttt{chunkLength} is a variable that is derived from data stored in a PNG file. The modified code allows up to two bytes past the end of the \texttt{in} array to be read. Since \texttt{in} is a pointer to memory provided by the consumer of the LodePNG library, it could be stored on either the stack or heap. IdSan was able to identify this error after only a few thousand executions of the program. We also fuzzed the same program with IdSan disabled for 1.5 million executions, and did not identify the same issue.

Not every change that we tried resulted in a detection from IdSan. We are unsure if this is an issue with IdSan, or whether the modification that we made did not introduce any memory safety issues. There appear to be redundant bounds checks in LodePNG, which might explain why some of the bounds checks that we tried to remove did not result in any apparent memory safety issues.

\section{Performance}
We analyzed the runtime overhead of IdSan on an unoptimized and an optimized build of \emph{LAME}. As input, we provided a 213KiB, mp3, 56 kbps, 44.1 kHz, Monaural 30-second file (mp3). The results can be seen in Table \ref{tbl:performance}. We list the execution time of the program running in unicornafl without IdSan hooked as a baseline, and the time with IdSan hooked to demonstrate the slowdown incurred. As expected the execution speed is reduced greatly by IdSan.

\begin{center}
    \begin{tabular}{|c|c|c|c|}
        \hline
        Build (\texttt{gcc} flag) & Without IdSan & With IdSan & Slowdown factor \\
        \hline
        Unoptimized (\texttt{-O0})  & 1.8s & 57.4s & ~31.8x \\
        Optimized (\texttt{-O3})    & 1.0s & 11.3s  & ~11.3x \\
        \hline
    \end{tabular}
    \label{tbl:performance}
\end{center}

In benchmarks performed by RetroWrite \cite{retrowrite}, the RetroWrite tool slows down execution by a factor of 1x-7x and AddressSanitizer with source code is able to achieve 1x-4x. Note however, that those benchmarks were done on an x86-64 architecture, without emulation layer and on different binaries.

This difference in performance between the two builds can be attributed to the amount of memory reads and writes that are done in the unoptimized build. Memory reads and writes incur a larger overhead because IdSan needs to copy the metadata from register metadata to the memory metadata, or vice versa. In the unoptimized build, every local variable is stored on the stack. For every update of a variable, there will be at least one memory read and one memory write in addition to the update itself, while in the optimized build the variable will likely be stored in a register, so no additional reads or writes are required. Because local variables are more likely to be pointers, none of the implemented fast paths for memory loads and stores of normal data can be applied. 

\section{Configuration}
IdSan was designed to be able to handle as many common patterns as possible without issues. It should be possible to run almost any standalone binary in the sanitizer with relatively little work. There are a number of configuration options available that can be enabled on a function-by-function basis for edge cases:

\begin{enumerate}
    \item Errors can be suppressed entirely for a function and all its callees, for example \texttt{malloc} and \texttt{free} are functions that usually walk the entire heap by casting pointers to different types. For such cases, memory safety violations can simply be ignored as they are intentional.
    \item Stack pointer identification can be delayed which can remove false positives in \texttt{va\_args} functions or functions with huge stack frames.
    \item Aligned reads up to 7 bytes beyond the end of a pointer can be allowed to support certain optimizations, which can be seen for example in implementations of \texttt{strchr}.
\end{enumerate}

\section{Limitations}
There are some scenarios where IdSan cannot effectively sanitize operations:

\begin{enumerate}
    \item Allocators (like \texttt{malloc} and \texttt{free}, as well as custom allocators) usually cannot be sanitized at all and memory errors need to be ignored in order to be able to run the program in the sanitizer.
    \item If the compiler optimizes variable accesses by directly adding offsets to different pointers, rather than go through either the \texttt{adrp} instruction or the \texttt{add sp, imm} pattern, the variables cannot be sanitized as separate objects and will instead need to be marked as a single bigger variable.
    \item Under less common scenarios, like the one discussed in Section \ref{invalid-mem-access} or when multiple computations on the stack pointer are done before it is used, additional configuration is needed to allow the sanitizer to sanitize effectively.
\end{enumerate}

\newgeometry{margin=1cm}
\begin{landscape}
\section{Comparisons with other sanitizers}
\begin{savenotes}
    We compare various memory sanitizers in table \ref{tbl:comparison}. Unfortunately some implementation details of MAI are unclear, for example the level and format of the human input required for sanitizing global variables is unknown, so we are not able to compare it with IdSan in more detail.
    \begin{center}
    \begin{table}[h!]
        \begin{tabular}{ |p{3cm}|p{3cm}|p{3cm}|p{3cm}|p{3cm}|p{3cm}|p{3cm}|p{3cm}| } 
            \hline
            & \multicolumn{3}{|c|}{Location-based} & \multicolumn{4}{|c|}{Identity-based} \\
            \cline{2-8}
                                & ASan                          & QAsan                         & RetroWrite                                & Safe-C                            & BBC       & MAI                   & IdSan \\
            \hline
            Source/binary?      & Both (source-only with afl)   & Binary                        & Binary \footnote{only PIE executables}    & Source                            & Source    & Binary                & Binary \\
            Platforms           & x86, x86\_64, arm, AArch64    & x86, x86-64, arm, AArch64     & x86-64                                    & Any                               & Any       & x86(-64)              & AArch64 \\
            Instrumentation via & compiler                      & qemu                          & static rewrite                            & semi-automated source translation & compiler  & unknown 
    \footnote{The authors state that they "plan to [..] migrate the implementation to Valgrind platform". The current implementation might be using a forked version of Valgrind, or possibly Intel Pin.}   & unicorn, ptrace \\
            \hline
            Stack?              & Source only                   & No                            & No                                        & Yes                               & Yes       & Yes                   & Yes \\
            Globals?            & Source only                   & No                            & No                                        & Yes                               & Yes       & Human input           & Human input \\
            Temporal violations?\footnote{For example: use-after-free bugs or use of uninitialized memory}
                                & Yes                           & Not implemented               & Yes                                       & Yes                               & Yes       & Yes                   & Not implemented \\
            Instrument dynamically generated 
            code?               & No                            & Yes                           & No                                        & No                                & No        & Yes                   & Not implemented \\
            \hline
        \end{tabular}
        \caption{A comparison of seven memory sanitizers.}
        \label{tbl:comparison}
    \end{table}
    \end{center}
\end{savenotes}
\end{landscape}
\restoregeometry
\chapter{Future work}

Future work on IdSan can be on improving precision (1-2), extending the sanitizer to handle more platforms or more violations (3-4), improving performance (5-6) and improving usability (7):

\begin{enumerate}
    \item We picked a very precise pointer metadata representation, allowing for arbitrary ranges of a value to represent different pointers. It is likely that in most cases such precision is not needed. The question "how much is enough?" still remains unanswered. Right before a memory access, the pointer information must always be reduced to a single $(\textit{base}, \textit{length})$ pair, which means that we only need to track enough information to make this reduction work. We suspect that storing information per-byte rather than per-bit and leaving out which specific range those bits represent should be precise enough, but we are unable to back this claim up with evidence.
    \item It might be possible to do more fine-grained sanitization of structures, like MAI \cite{mai} attempts to do. However, there appear to be a number of issues with memory copying and compiler optimizations that would need to be solved in order for this to work.
    
    \item IdSan currently only sanitizes for spatial memory safety violations. However, this could be extended to also include temporal safety violations.
    \item It should be possible to extend IdSan to support other architectures like x86-64 and 32-bit ARM with relatively little work.
    \item afl uses forking to spawn new processes. Right now, a new process is forked before the program in unicornafl is started. Fuzzing could be sped up by forking after the program running inside unicornafl has initialized. This would of course require cooperation from the program running inside unicornafl, making the fuzzing process more complicated. However, we believe this trade-off to be worthwhile, as it would allow programs that spend a large part of their execution time initializing (like the LAME library we tried to fuzz) to be fuzzed much quicker.
    
    \item It might be possible to improve performance by running the sanitizer on multiple threads. Because of pipelining in processors, compilers are incentivized to spread out dependent instructions, and interleave them with other instructions so that the processor pipeline does not stall. In the same way, it might also be possible to run the sanitizer on multiple threads and dispatch the sanitizer operation to the correct thread depending on which registers or memory it depends on.
    
    \item Currently, all information on the stack frames is input manually. A tool to automate (part) of this process could help make our approach more viable.
\end{enumerate}

We also ran into some more general issues:

\begin{enumerate}
    \item Comparisons with other tools are currently quite limited because IdSan only supports AArch64, while most other tools aim for x86-64 instead. This makes directly comparing tools difficult. As more tools gain support for AArch64, better comparisons can be made.
    \item There exists no good code sample set that attempts to cover a large range of vulnerabilities. This makes writing and testing a sanitizer more difficult, as everyone has to develop their own set of test cases. Development of a well-tested set of code samples that covers a large range of 
\end{enumerate}
\chapter{Conclusion}
Initially, the goal of this research was to simply find a way to sanitize binary programs while fuzzing them. With the introduction of QAsan a few months before starting, we shifted our focus to finding a way to sanitize stack and global variables on binaries, as this problem has not been explored much. Most of the recent work \cite{qasan} \cite{retrowrite} \cite{bbc} \cite{paricheck} uses a location-based approach, which we assume is likely because of performance advantages. It is also possible that sanitizing of just the dynamic allocations is considered "good enough" in practice.

We designed and implemented a sanitizer for binary programs without source code. Sanitizing stack- and global variables using a location-based approach is much more difficult, if not impossible, because of the need to recompile the code as described in section \ref{locsan-globs}. Therefore, IdSan is identity-based rather than location-based, which allows us to sanitize stack- and global variables in addition to dynamically allocated memory.

Because of the limited literature available \cite{mai} on the design of identity-based sanitizers for binaries, we opted for a generic approach that favors development time over performance. The metadata that we are storing is also likely to be an overestimation of the metadata that is really needed to effectively and accurately sanitize a binary. We list a number of issues that need to be taken into account when building an identity-based sanitizer in Section \ref{design-considerations}.

Compared to MAI \cite{mai}, which attempts to automatically infer memory structure at runtime, we only guess some sane defaults in case of missing information. Most information is loaded from a file instead. We are unable to achieve the same level of fine-grainedness as claimed by MAI \cite{mai}, because this will lead to false-positives as soon as a datastructure is copied to another memory location (see section \ref{memcpy-issues}). Our approach does not have some of the limitations described by the authors of MAI:

\begin{enumerate}
    \item MAI is unable to handle NULL-terminated strings, because it may see the NULL byte as a separate object from the text before it. IdSan does not have this issue, because it does not infer variable sizes and instead relies on manual input.
    \item Similarly, cases where the compiler optimizes two pointer additions like $p_1 = p_0 + 5$ and $p_2 = p_1 + 10$ into $p_2 = p_0 + 15$ trip up MAI, but are handled without issue by IdSan.
    \item MAI may produce false positives on union fields, while our tool is able to handle these without issues.
\end{enumerate}

The trade-off here is that while IdSan might produce fewer false positives, it also produces fewer true positives. The exact details of this trade-off are unknown, as we do not have access to MAI's source code.

Adding a new instruction set to the sanitizer should be relatively little work, assuming that it maps onto the sanitizer operations. Expanding to for example RISC should be relatively easy, while expanding to x86 and its various extensions is likely to be slightly more difficult.

\printbibliography

\appendix
% \chapter{A chapter}

\end{document}